\pdfoutput=1
\RequirePackage{ifpdf}
\ifpdf % We are running pdfTeX in pdf mode
\documentclass[pdftex]{sigma}
\else
\documentclass{sigma}
\fi

\begin{document}

\allowdisplaybreaks

\renewcommand{\thefootnote}{$\star$}

\renewcommand{\PaperNumber}{012}

\FirstPageHeading

\ShortArticleName{Specialized Orthonormal Frames and~Embedding}

\ArticleName{Specialized Orthonormal Frames and~Embedding\footnote{This
paper is a~contribution to the Special Issue ``Symmetries of Dif\/ferential Equations: Frames,
Invariants and~Applications''.
The full collection is available
at
\href{http://www.emis.de/journals/SIGMA/SDE2012.html}{http://www.emis.de/journals/SIGMA/SDE2012.html}}}

\Author{Frank B.~ESTABROOK}

\AuthorNameForHeading{F.B.~Estabrook}

\Address{Jet Propulsion Laboratory, California Institute of Technology,\\
4800 Oak Grove Drive, Pasadena, CA 91109 USA}
\Email{\href{mailto:frank.b.estabrook@jpl.nasa.gov}{frank.b.estabrook@jpl.nasa.gov}}

\ArticleDates{Received October 09, 2012, in f\/inal form February 12, 2013; Published online February 15, 2013}

\Abstract{We discuss some specializations of the frames of f\/lat orthonormal frame bundles over
geometries of indef\/inite signature, and~the resulting symmetries of families of embedded
Riemannian or pseudo-Riemannian geometries.
The specializations are closed sets of linear constraints on the connection 1-forms of the framing.
The embeddings can be isometric, as in minimal surfaces or Regge--Teitelboim gravity, or
torsion-free, as in Einstein vacuum gravity.
Involutive exterior dif\/ferential systems are given, and~their Cartan character tables
calculated to express the well-posedness of the underlying partial dif\/ferential embedding
and~specialization equations.}


\Classification{83C20; 57R40; 58A15}

\renewcommand{\thefootnote}{\arabic{footnote}}
\setcounter{footnote}{0}

\section{Introduction}

The class $C$ of a~Riemannian space is the minimum number of extra dimensions required for it to be
locally embeddable in a~f\/lat space (various signatures possible, which we ignore for the moment;
often we write repeated indices as ``covariant", but proper signs must be inserted in such
summations).
Also our discussions are all local, with no global existence implied for solutions of the
equivalent partial dif\/ferential equation.
So class $C$ 4-metrics are those embeddable in $4+C$ f\/lat dimensions, $C\leq6$, and~the embedding
of their associated orthonormal frame bundles (Cartan moving frame formalism, basis forms
$\theta_{i}$, connection forms $\omega_{ij}=-\omega_{ji}$, $i=1,\dots,4$) brings in $C+4C+C(C-1)/2$
additional orthonormal basis forms $\theta_{A}$ and~connection forms $\omega_{iA}=-\omega_{Ai}$
and~$\omega_{AB}=-\omega_{BA}$, $A=1,\dots,C$.
Taking $\mu$, $\nu$, etc.\
to span the combined range of indices $1,\dots,4+C$ of the embedding space, the 1-forms $\theta_{\mu}$
and~$\omega_{\mu\nu}=-\omega_{\nu\mu}$ satisfy the Cartan--Maurer structure equations of a~f\/lat
orthonormal frame bundle ${\rm ASO}(4+C)$.\footnote{Thus Einstein vacuum general relativity can be set as
an exterior dif\/ferential system on the group space ${\rm ASO}(10)$, of dimension 55.
This is arguably an elegant improvement over the more customary setting of this f\/ield theory in
terms of coordinate components, which means for a~f\/irst order set of PDE's setting the EDS on the
second frame bundle over a~four dimensional base, of dimensionality $4+16+40=60$.
The f\/irst EDS has 21 Cauchy characteristic vectors so in principle 21 variables can be
eliminated, leaving a~system of partial dif\/ferential equations in 34 variables, 4 in involution,
independent, and~30 dependent.
The second setting, when using orthonormal tetrad frames, as in the dyadic and~Newman--Penrose
formalisms, only reduces to 44 variables but has 10 degrees of gauge freedom together with 10
conservation laws~\cite{FEMath}.}

Isometric embedding requires the $\theta_A$ to vanish upon pullback, and~together with their
exterior derivatives the 2-forms $\omega_{Ai}\wedge\theta_i$, and~dynamic 4-forms derived from the
Hilbert Lagrangian, $\omega_{iA}\wedge R_{jk}\wedge\theta_{l}\epsilon^{ijkl}$, they generate well
posed exterior dif\/ferential systems (EDS) ($R_{ij}$ are the induced Riemann 2-forms
$d\omega_{ij}+\omega_{ik}\wedge\omega^{k}_{j}$)~\cite{Ivey}.
These isometric EDS's for solutions of Regge--Teitelboim (RT) gravity, of various classes, have
been expounded in a~recent paper~\cite{Esta1}.

Among the solutions of RT theory are those that additionally annul the compatible Ricci 3-forms
$\omega_{iA}\wedge\omega_{Aj}\wedge\theta_{k}\epsilon^{ijkl}$.
This more restricted vacuum Einstein theory can conveniently be directly set as a~(not strictly
isometric) embedding EDS on the orthonormal frame bundle over 10 space, requiring only the four
torsion 2-forms $\omega_{iA}\wedge\theta_{A}$ to pull back to vanish, together with their exterior
derivatives and~the dynamic Ricci 3-forms.
A number of vacuum solutions, as well as other physically interesting non-vacuum metrics, have been
classif\/ied both by embedding class and~motion group, or metric symmetry~\cite{Exact}.
Here we treat both EDS's for embedded families of solutions of RT theory with special symmetry
and~class ${\leq}6$, and~new EDS for families of solutions of Einstein theory with special symmetry
in class~6.

In~\cite{Esta1} an explicit six dimensional framing for embedding static spherically symmetric
(SSS) metrics was found, and~solutions of Regge--Teitelboim theory reduced to quadrature, using
a~coordinate imbedding due to Ikeda (other types of class~2 SSS coordinate embeddings exist; with
two new ones, all possible have since been given by Paston and~Sheykin~\cite{Past}).
The Schwarzschild vacuum solution of orthodox general relativity was a~particular one of these.
The connection forms there derived were not explicitly given; we write them below in Section~\ref{section2}.
We extract the algebraic specialization of the ${\rm O}(6)$ frame f\/ibers they imply.
The specialization is a~set of self closed linear relations between the $\omega_{\mu\nu}$.
A closely related specialization is also presented and~Cartan characters of embedding RT EDS's
using these specialized frames are calculated, showing only the second of them to be well posed.
We take it to be the prototype for a~new method of frame specialization that can be used for
formulating EDS's for families of solutions having specialized metric symmetries.

In Section~\ref{section3} we outline this technique for specializing the f\/ibers of ${\rm ASO}(4+C)$ frame bundles,
and~present calculated Cartan character tables of EDS's for isometric embedding of RT theory in
them, showing them to be well posed and~involutory.
Section~\ref{section4} brief\/ly expounds an illustrative toy problem, isometrically embedding a~two
dimensional variational problem in a~frame specialized f\/lat 3-space.
In Section~\ref{section5} we give results for four specializations of the f\/ibers of ${\rm ASO}(6,4)$, the
orthonormal frame bundle over f\/lat ten-dimensional space.
With torsion 2-forms and~Ricci 3-forms, we f\/ind EDS's for families of Einstein vacuum solutions
that are well posed and~involutory.

To summarize, we have calculated the Cartan characters, and~well posedness, of a~number of EDS for
embedded 4-spaces in specialized higher dimension f\/lat orthonormal frame bundles, and~we
conjecture that their solutions have intrinsic symmetry properties such as Riemannian submersions,
isometries and~Killing tensors.
The specializations we consider are sets of self closed linear relations between the
$\omega_{\mu\nu}$.
Families of metrics with symmetries have heretofore been dif\/f\/icult to treat with Cartan moving
frame or tetrad formalism~\cite{Esta2}.

\section{A static spherically symmetric embedding}\label{section2}

In~\cite{Esta1}, treating class 2 static spherically symmetrical spacetimes, which all have four isometries, they
were imbedded in a~f\/lat 6-space coordinatized by $r$, $\theta$, $\phi$, $t$, $z_5$, $z_6$, signature
$(-1,-1,-1, 1,1,-1)$; The 6-space was framed with a~family of orthonormal hexad bases adapted to the
(already known) metrics of the embedded space-times: This framing could have been even more
general, and~its connection forms were not there written out, but they allow us to induce the
specialization of connection forms that succinctly expresses the desired symmetry; the fra\-ming~is
\begin{gather*}
\theta_1=\Bigl(dr+\big(dz^5+dz^6\big)c'[r]-2c'[r]t(2c'[r]t dr+2c[r]dt)
\\
\hphantom{\theta_1=}{}
-\big(dz^5-dz^6\big)\big(g'[r]+c'[r]t^2\big)\Bigr)\Big/\sqrt{1+4c'[r]g'[r]},
\\
\theta_2=r d\theta,
\qquad
\theta_3=r\sin[\theta]d\phi,
\qquad
\theta_4=t dz^5-t dz^6+2c[r]dt+2c'[r]t dr,
\\
\theta_5=\Bigl(dr+dz^5\big(g'[r]-c'[r]t^2+c'[r]+1/(2c'[r])\big)-2t c'[r](2c'[r]t dr+2c[r]dt)
\\
\qquad
-dz^6\big(g'[r]-c'[r]t^2-c'[r]+1/(2c'[r])\big)\Bigr)\Big/\sqrt{1+4c'[r]g'[r]},
\\
\theta_6=dr+dz^5/(2c'[r])-dz^6/(2c'[r]),
\end{gather*}
$c[r]$ and~$g'[r]$ are arbitrary functions.
The f\/lat metric is
\begin{gather*}
-\theta_1\theta_1-\theta_2\theta_2-\theta_3\theta_3+\theta_4\theta_4+\theta_5\theta_5-\theta_6\theta_6,
\end{gather*}
and the embedding is isometric, $\theta_5=0$, $\theta_6=0$.

The exterior derivatives of the above, expanded on them, give an explicit set of specialized the 15
connection 1-forms~$\omega_{\mu\nu}$
\begin{gather*}
\omega_{12}=\frac{\theta_2}{r\sqrt{1+4c'[r]g'[r]}},
\qquad
\omega_{13}=\frac{\theta_3}{r\sqrt{1+4c'[r]g'[r]}},
\qquad
\omega_{14}=\frac{c'[r]\left(\theta_4-2t\theta_6c'[r]\right)}{c[r]\sqrt{1+4c'[r]g'[r]}},
\\
\omega_{15}=
\Big(\theta_1-\theta_5+\theta_6\mathop{\sqrt{1+4c'[r]g'[r]}}\Big)\Big(-\bigl(1+2c'[r]g'[r]\bigr)c''[r]
\\
\hphantom{\omega_{15}=}{}
+2c'[r]^2g''[r]\Big)\Big/\Bigl(c'[r]\bigl(1+4c'[r]g'[r]\bigr)^{3/2}\Bigr),
\\
\omega_{16}=-\Big(\bigl(\theta_1-\theta_5\bigr)\sqrt{1+4c'[r]g'[r]}+
\theta_6\bigl(1+4c'[r]g'[r]\bigr)\Big)c''[r]\Big/\Bigl(c'[r]\bigl(1+4c'[r]g'[r]\bigr)^{3/2}\Bigr),
\\
\omega_{23}=\frac{\text{Cot}[\theta]\theta_3}{r},
\qquad
\omega_{24}=0,
\qquad
\omega_{25}=-\frac{\theta_2}{r\sqrt{1+4c'[r]g'[r]}},
\qquad
\omega_{26}=-\frac{\theta_2}{r},
\qquad
\omega_{34}=0,
\\
\omega_{35}=-\frac{\theta_3}{r\sqrt{1+4c'[r]g'[r]}},
\qquad
\omega_{36}=-\frac{\theta_3}{r},
\qquad
\omega_{45}=\frac{c'[r]\left(-\theta_4+2t\theta_6c'[r]\right)}{c[r]\sqrt{1+4c'[r]g'[r]}},
\qquad
\omega_{46}=0,
\\
\omega_{56}=-\Big(\bigl(\theta_1-\theta_5\bigr)\sqrt{1+4c'[r]g'[r]}+
\theta_6\bigl(1+4c'[r]g'[r]\bigr)\Big)c''[r]\Big/\Bigl(c'[r]\bigl(1+4c'[r]g'[r]\bigr)^{3/2}\Bigr).
\end{gather*}

To generalize from this coordinate presentation, we note from it that the algebraic frame f\/iber
specialization for this SSS embedding is apparently generated by the annulling of seven 1-forms
$\omega_{12}+\omega_{25}$, $\omega_{13}+\omega_{35}$, $\omega_{14}+\omega_{45}$,
$\omega_{16}+\omega_{65}$, $\omega_{24}$, $\omega_{34}$, $\omega_{64}$.
It can be checked that: exterior derivatives of these connection 1-forms all vanish by virtue of
the 1-forms themselves.
Otherwise said, they are self-closed, or a~completely integral Pfaf\/f\/ian system (c.i.p.s.).
Let us call them a~set of {\it{symmetry forms}}.

We might consider an isometric embedding EDS with this frame specialization, set on the 21
dimensional orthonormal frame bundle over six dimensions $\theta_{i}$, $i = 1,\dots,4$ and~$\theta_{A}$,
$A = 5,6$.
The EDS is generated by the nine 1-forms ($\theta_{A}$,  {\it symmetry forms}), together with
their closure 2-forms.
Taking the signature $-1,-1,-1,1,1,-1$ we calculate the EDS to have Cartan character table
$21(9,2,2,2)4$ and~to be in involution with respect to the~$\theta_{i}$.
Since $\omega_{23}$ does not appear in the generators there is one Cauchy Characteristic vector.
There is one degree of gauge freedom: $s_{4}=1$.
If we further add to this EDS the two dynamic 4-forms of Regge--Teitelboim gravity, however, we no
longer have a~well posed system: all four~$\theta_{i}$ are then not in involution.
So this isometric embedding EDS is apparently not exactly equivalent to the coordinate presentation
of~\cite{Esta1}.
The coordinate presentation has of course further specialized the framing (only two arbitrary
functions of one metric variable come in), and~we are in this present work concerned with f\/inding
well posed involutory families of symmetric solutions.

A possible resolution discussed in Section~\ref{section3} is to consider an EDS generated by the closely
related c.i.p.s.\
$\omega_{12}+\omega_{25}$, $\omega_{13}+\omega_{35}$, $\omega_{14}+\omega_{45}$,
$\omega_{16}+\omega_{65}$, $\omega_{42}+\omega_{26}$, $\omega_{43}+\omega_{36}$.
This set of six symmetry forms leads to a~well posed class 2 EDS for RT theory, with character
table $21(8,2,2,4)4$, in involution with respect to the $\theta_{i}$ and~has one degree of gauge
freedom $s_{4}=1$.

\section{A systematic specialization of frames in embedding spaces}\label{section3}

\looseness=-1
The f\/irst four of the symmetry forms for Schwarzschild embedding found above are a~c.i.p.s.\ in
their own right: $\omega_{12}+\omega_{25}$, $\omega_{13}+\omega_{35}$, $\omega_{14}+\omega_{45}$,
$\omega_{16}+\omega_{65}$.
The exterior derivative of any of these pulls back to vanish modulo the set.
This is also true if all four are written with minus signs between the two terms.
The index 1, or in general $i$, is from the set of imbedded forms, and~the index 5 or in general
$A$ is from the coset of embedding dimensions.
The pulled back form $\omega_{15}$, or in general $\omega_{iA}$, is exact modulo the set.
Remarkably, this construction of a~c.i.p.s.\
works for all dimensions and~signatures, \emph{so long as the signatures of $i$ and~$A$ are of
opposite sign}! We remember the Codazzi equation $d\omega_{iA}+\omega_{i\mu}\wedge\omega_{\mu A}=0$
and~notice that all the quadratic terms vanish modulo these c.i.p.s.\
This is the most basic construction of specializing symmetry forms we have found.

Further systematic addition of these specializing sets of symmetry forms can go on as long as more
pairs, say $j$, $B$, etc., of opposite signature are available (omitting previous $i$, $A$, etc.
pairs from each successive summation), and~a~number of Cartan characteristic table calculations have
shown that well posed RT (isometric) EDS's are always found.
Moreover, we f\/ind this again in the class 6 case for Einstein vacuum (torsion-free embedding).
In Section~\ref{section5} we explicitly write these four c.i.p.s., having numbers of generators $8$, $8+6=14$,
$8+6+4=18$ and~$8+6+4+2=20$ respectively.

  Our systematic construction of these c.i.p.s.\
for specializing frame curvature forms, in
\linebreak ${\rm ASO}(n,m)$, appears to be new, but a~number of other
c.i.p.s.\
for specialized frames are of course well known and~important.
When a~c.i.p.s.\
is imposed on the f\/iber algebra ${\rm O}(n,m)$ the remaining moving frame forms are those of
a~subbundle, which can be labeled by its f\/iber algebra and~base dimension.
For example the case of a~subbundle ${\rm ASU}(m)$ of ${\rm ASO}(m,m)$, and~further isometric embedding of
${\rm ASO}(m)$ in that, is explicitly treated as an EDS for special Lagrangian calibration in~\cite[p.~201]{Ivey}
(the Cartan character table given there should read $s_2=n-1$).
Various other interesting manifolds discussed there similarly can be understood as discovery of
c.i.p.s.\
for subbundles of orthonormal frame bundles ${\rm ASO}(m,n)$; taken together with isometric embedding
derived from minimizing calibration forms, EDS are found for associative, coassociative and~Cayley
geometries.
We should also stress that, in distinction to previous such work (e.g., Calabi--Yau manifolds), our new specializations have so far only been been treated locally, through calculation of
Cartan characters of the various embedding EDS.

\section{The simplest case: a~specialized embedding\\ of minimal 2-spaces in f\/lat 2+1-space}\label{section4}

With moving frame bases $\theta_{i}$, $i=1,2$, and~$\theta_{A}$, $A=3$, signature $(1,1,-1)$ we begin with
an EDS generated by $\theta_{3}$, its exterior derivative
$\omega_{13}\wedge\theta_{1}+\omega_{23}\wedge\theta_{2}$, and~the new specialization which in
three dimensions is just the single form $\omega_{12}-\omega_{13}$.
The character table is $6(2,1)2$ with one degree of gauge freedom, well posed and~involutory on
$\theta_{1}$ and~$\theta_{2}$.
As RT theory does not exist in this low dimension we then consider extremal surfaces, with
Lagrangian $\Lambda=\theta_{1}\wedge\theta_{2}$, whose variation generates the EDS to be the
isometric embedding form $\theta_{3}$ together with the dynamic 2-form
$\omega_{13}\wedge\theta_{2}-\omega_{23}\wedge\theta_{1}$.
This EDS for specialized minimal surfaces remains well posed, involutory on $\theta_{1}$
and~$\theta_{2}$, with table $6(2,2)2$.

Integration, introduction of scalar variables to explicitly f\/ind the PDE's, begins with the
closed 1-form $\omega_{23}$, leading to a~coordinate we call $u$.
Sequentially from the structure (moving frame) equations and~2-forms in the EDS we f\/ind further
closed 1-forms, and~can introduce variab\-les~$v$,~$x$ and~$y$.
The result is {\samepage
\begin{gather*}
\omega_{23}=-u^{-1}du,
\qquad
\omega_{13}=u^{-1}dv,
\qquad
\theta_{2}=u dx,
\qquad
\theta_{1}=u dy,
\end{gather*}
and the two 2-forms require
$
u_{x}-v_{y}=u_{y}+v_{x}=0$. }

The induced intrinsic 2-metric on any of the family of solutions is then seen to be
\begin{gather*}
ds^2=\theta_{1}\theta_{1}+\theta_{2}\theta_{2}=u^2\big(dx^2+dy^2\big),
\end{gather*}
where \looseness=-1 $u(x,y)$ satisf\/ies Laplace's equation.
The Gaussian curvature is $(u_x^2+u_y^2)/u^4$.
We have not yet been able to characterize a~common geometric property of this family of embedded
minimal surfaces, resulting from various solutions of Laplace's equation; but it may be relevant
that at least the plane, point particle and~dipole solutions all lead to metrics that support
a~Killing vector f\/ield.

\vspace{-1mm}

\section{Application to Einstein gravity}\label{section5}

Not surprisingly the symmetry specializations we have found to work with Regge--Teitelboim gravity
of various classes appear to work also with Einstein gravity, but \emph{only} in the class~6 case,
embedding in f\/lat 10-dimensional space with the required signature.
Then indeed we f\/ind specialized frame bundles for setting EDS's for families of solutions of
vacuum general relativity.
This means we drop the strict isometric embedding 1-forms and~return to the EDS (equally generated
from the Hilbert Lagrangian) of four torsion 2-forms and~four Ricci 3-forms.

Let us number bases $i=1,2,3,4$ and~$A = 5,6,7,8,9,0$, use signature
$(-1,-1,-1,1,1,1,1$, $-1, 1,1)$
and~base our f\/irst specialization on indices 1 and~5.
Take the symmetry forms to be the c.i.p.s.\
$(\omega_{12}+\omega_{25}$, $\omega_{13}+\omega_{35}$, $\omega_{14}+\omega_{45}$, $\omega_{16}-\omega_{56}$,
$\omega_{17}-\omega_{57}$, $\omega_{18}-\omega_{58}$, $\omega_{19}-\omega_{59}$, $\omega_{10}-\omega_{50})$.
The EDS is generated by these eight 1-forms, four torsion 2-forms $\omega_{iA}\wedge\theta_{A}$,
their exterior derivatives, and~the dynamic Ricci 3-forms
$\omega_{iA}\wedge\omega_{Aj}\wedge\theta_{k}\epsilon^{ijkl}$.
Explicit calculation of the Cartan characters of this EDS shows it to be well posed and~involutory
on the $\theta_{i}$: $55(8,4,12,14)4$ with three degrees of gauge freedom.
There are 10 Cauchy characteristic vectors
($\omega_{67}$, $\omega_{68}$, $\omega_{69}$, $\omega_{60}$, $\omega_{78}$, $\omega_{79}$, $\omega_{70}$,
$\omega_{89}$, $\omega_{80}$, $\omega_{90}$ do not appear in the generators of the EDS).

The second specialization adds the additional six symmetry forms based on indices 2 and~6:
$(\omega_{23}+\omega_{36}$, $\omega_{24}+\omega_{46}$, $\omega_{27}-\omega_{67}$, $\omega_{28}-\omega_{68}$,
$\omega_{29}-\omega_{69}$, $\omega_{20}-\omega_{60})$.
Again well posed; $55(14,4,12,14)4$, one degrees of gauge freedom and~six CC vectors
($\omega_{78}$, $\omega_{79}$, $\omega_{70}$, $\omega_{89}$, $\omega_{80}$, $\omega_{90}$ do not appear).

The third specialization adds four symmetry forms based on 3 and~7,
$55(18,4,12,14)$ with 2~CC's.
The fourth specialization adds yet two more, based on 4 and~8, so 20 in all.
This results in $55(20,4,12,14)4$, with one remaining gauge freedom.

\medskip
{\bf Acknowledgements.}
I thank the JPL Of\/f\/ice of the Chief Scientist for a~visiting appointment during which this
research was carried out, and~the Science Division for hospitality.
My colleagues John W.~Armstrong, Curt Cutler, Massimo Tinto and~Michele Vallisneri gave constant
stimulus and~support.

\vspace{-1mm}

\pdfbookmark[1]{References}{ref}
\LastPageEnding

\end{document}